\documentclass[preprint,showpacs,pre,12pt]{revtex4-1}
\usepackage{amsfonts}
\usepackage{amssymb}
\usepackage{amsmath}
\usepackage{graphicx}

\begin{document}
\title{Modification of the Porter-Thomas distribution by  rank-one interaction}
\author{E. Bogomolny}
\affiliation{CNRS, Universit\'e Paris-Sud, UMR 8626\\
Laboratoire de Physique Th\'eorique et Mod\`eles Statistiques, 91405 Orsay,
France}
\date{\today}

\begin{abstract}
The Porter-Thomas (PT) distribution of resonance widths is one of the oldest and simplest applications of statistical ideas in nuclear physics. Previous experimental data confirmed it quite well but recent and more careful investigations show clear deviations from this distribution.  To explain these discrepancies the authors of Ref.~\cite{volya}, [PRL  \textbf{115}, 052501 (2015)],  argued that to get  a realistic model of nuclear resonances is not enough to consider one of the standard random matrix ensembles which leads immediately to the PT distribution but  it is necessary to add a rank-one interaction which couples resonances to decay channels. The purpose of the paper is to solve  this model analytically and to find explicitly the  modifications of the PT distribution due to such interaction. Resulting formulae are simple, in a good agreement with numerics, and could explain experimental results.
\end{abstract}
\maketitle

\section{Introduction}

Random matrix (RM) theory has undeniable success in describing nuclear physics data,  in particular statistical properties of nuclear resonances and the distribution of their widths (see e.g. reviews \cite{porter}--\cite{fyodorov} and references  therein). One of the simplest and widely used  RM predictions is the statement that resonance widths are distributed as modulus square of RM eigenfunctions.   For large dimensional invariant ensembles the latter are described by the Gaussian  distribution which leads to the famous  Porter-Thomas (PT) law \cite{PorterThomas}  
 \begin{equation}
P_1(x)=\frac{1}{\sqrt{2\pi l x}}\exp\left ( -\frac{x}{2l}\right ).
\label{pt_1}
\end{equation} 
In nuclear physics $x$ is reduced resonance width and in RM theory  $x=N|\Psi |^2$ where $\Psi$ is any eigenfunction component and $N$  is  the matrix dimension.  This expression is valid for 
time-symmetric systems (denoted below by index $\beta=1$ or GOE). For time-non-invariant systems  (denoted by index $\beta=2$ or GUE) real and imaginary parts of RM eigenfunctions are independent Gaussian random variables which gives 
\begin{equation}
P_2(x)=\frac{1}{l} \exp\left ( -\frac{x}{l}\right ).
\label{pt_2}
\end{equation} 
Constant $l$ equals the mean value of $x$. Standard choice is $\langle x \rangle=1$ and $l=1$.

In RM theory the  PT law is a theorem for invariant ensembles in the limit $N\to\infty$ (see e.g. \cite{mehta}). For physical problems like nuclear resonances its applicability is not guaranteed and requires experimental verification. Older experiments (cf. \cite{porter}, \cite{weidenmuller}) were in reasonable good agreement with this law. Nevertheless, recent experimental results and more careful treatment  of old results demonstrate a clear disagreement with  the Thomas-Porter distribution \cite{koehler_1}-\cite{koehler_3}.  As RM theory is one of  cornerstones of quantum chaos in nuclear physics, it is important to understand the origin of the discrepancy.   Different scenarios had been proposed so far (see \cite{hans}-\cite{volya} among others). 

After a careful analysis the authors of Ref.~\cite{volya} came to the conclusion  that a realistic model of nuclear $s$-wave resonances should include in addition to RM term a rank-one interaction which couples resonances to decay channels and  they argued that the effective Hamiltonian matrix can be chosen in  the form
\begin{equation}
M_{ij}=G_{ij}^{(\beta)}+Z\delta_{i1}\delta_{j1}\ .
\label{M_G}
\end{equation}
Here $G_{ij}^{(\beta)}$ is either a $N\times N$ real symmetric random matrix  ($\beta=1$) or a complex Hermitian one ($\beta=2$) with the Gaussian distribution
\begin{equation}
P(G_{ij})\sim \exp \left (-\frac{\beta}{4\sigma^2}\mathrm{Tr}\, (G\,G^{\dag})\right ).
\label{G_beta}
\end{equation}
To get a nontrivial limit  it is assumed that the ratio of the coupling constant $Z$ to the mean level density 
\begin{equation}
\kappa=\frac{Z}{\sigma \sqrt{N}}
\label{kappa}
\end{equation}
remains constant when $N\to\infty$.

Let $E_{\alpha}$  and $\Psi_i(\alpha)$ be eigenvalues and corresponding eigenvectors of matrix $M_{ij}$.
In Ref.~\cite{volya} it was noted that for $\beta=1$ the distribution of the quantity
\begin{equation}
x_{\alpha}=N|\Psi_1(\alpha)|^2
\label{x_alpha}
\end{equation}
does deviate from the PT law. But this conclusion was based  only on numerical calculations and no clear physical picture had emerged. 

The purpose of this work is to demonstrate that the eigenfunction distribution of  matrix \eqref{M_G} can be found analytically for large $N$. The main result of the paper is that random variable $x$ given by Eq.~\eqref{x_alpha} has the same functional form as the PT  distribution (Eqs.~\eqref{pt_1} and \eqref{pt_2}) but with important difference that $l$  (and, consequently, the mean value of $x$) in these expressions is not an universal constant but a certain function of state energy $E$ and dimensionless coupling constant $\kappa$ (see Eq.~\eqref{finalPT}). 

The plan of the paper is the following. In Section~\ref{rank_one} general  formulae useful for rank-one perturbation are briefly discussed for completeness. Section~\ref{probability} is devoted to the derivation of exact joint distribution of eigenvalues and eigenfunctions of matrix $M$ in Eq.~\eqref{M_G}.  It is demonstrated that for any rank-one perturbation such  distribution equals the unperturbed joint distribution of invariant matrix $G$ without the confinement term  \eqref{G_beta}. It is the change of the confinement potential which mixes eigenvalues and eigenfunctions and makes the problem rotational non-invariant. Section~\ref{asymptotic} contains  the determination of  eigenfunction distributions in the limit $N\to\infty$. The cases $\kappa^2<1$ and $\kappa^2>1$  are treated  separately as in the latter a collective state appears whose eigenfunction square  for large $N$ has a macroscopic (i.e. independent on $N$) value.  In the both cases the main  modification of the Thomas-Porter distribution due to the rank-one interaction consists in appearance of energy depended $l(E)$ in Eqs.~\eqref{pt_1} and \eqref{pt_2}. Such simple  Gaussian-like character of the resulting distribution is valid only for eigenfunctions  with energies lying in a small window. The distribution of eigenfunctions in  a large window is not Gaussian and is calculated in Section~\ref{large}. Section~\ref{conclusion} summarises the obtained results  and shortly discusses possible experimental applications. 

\section{Rank-one perturbation}\label{rank_one}

Let two Hermitian matrices $G$ and $M$  be related by a rank-one term
\begin{equation}
M_{ij}=G_{ij}+  v_i^{*}v_j  .
\label{def}
\end{equation}
To describe matrix \eqref{M_G}
 \begin{equation}
 v_j=\sqrt{Z} (1,0,\ldots,0).
 \label{v}
 \end{equation}
 Eigenvalues and eigenfunctions of these two matrices are ($\alpha=1,\ldots,N $)
\begin{equation}
\sum_{j=1}^N G_{ij}\Phi_j(\alpha)=e_{\alpha} \Phi_i(\alpha) , \qquad
\sum_{j=1}^N M_{ij}\Psi_j(\alpha)=E_{\alpha} \Psi_i(\alpha) .\label{M}
\end{equation}
All eigenfunctions are assumed to be orthogonal.

Eigenfunctions of new matrix $M$ can be expanded into a series of  eigenfunctions of matrix $G$
\begin{equation}
\Psi_j(\alpha)=\sum_{\beta=1}^N C_{\alpha  \beta}\Phi_j(\beta),\qquad \Phi_j(\alpha)=\sum_{\beta=1}^N C_{\alpha  \beta}^{-1}\Psi_j(\beta).
\end{equation}
Substituting this expansions to Eq.~\eqref{M} one gets 
\begin{equation}
C_{\alpha\, \beta}=\frac{a_{\alpha} b_{\beta}^*}{E_{\alpha}-e_{\beta}},\qquad 
b_{\beta}=\sum_{j=1}^N v_j \Phi_j(\beta),\quad a_{\alpha}=\sum_{\beta} C_{\alpha \beta}b_{\beta} .
\label{a_b}
\end{equation}
The last relation gives the quantisation condition 
\begin{equation}
 \sum_{\beta}\frac{|b_{\beta}|^2}{E_{\alpha}-e_{\beta}}=1.
 \label{quantisation_1}
\end{equation}
It is plain  that eigenvalues $E_{\alpha}$ and $e_{\alpha}$ are intertwined. 

Exactly in the same way one obtains complementary relations 
\begin{equation}
C_{\alpha\, \beta}^{-1}= \frac{\tilde{a}_{\alpha} \tilde{b}_{\beta}^*}{E_{\beta}- e_{\alpha}}\qquad 
\tilde{b}_{\beta}=\sum_{j=1}^N v_j \Psi_j(\beta),\qquad \tilde{a}_{\alpha}=\sum_{\beta} C_{\alpha \beta}^{-1}\tilde{b}_{\beta}^*,\qquad \sum_{\beta}\frac{|\tilde{b}_{\beta}|^2}{e_{\alpha}-E_{\beta}}=-1 .
\label{tilde_a_b}
\end{equation}
From Eqs.~\eqref{tilde_a_b} and \eqref{a_b} it follows that  $\tilde{b}_{\beta}=a_{\alpha}$ and $\tilde{a}_{\alpha}=b_{\alpha}$.

The above quantisation conditions can be solved for the numerators. Using the Cauchy determinant formula or contour integration as in Appendix of Ref.~\cite{integrable}  one concludes  that 
\begin{equation}
|b_{\alpha}|^2=\dfrac{\prod_{\gamma}(E_{\gamma}-e_{\alpha})}{\prod_{\gamma\neq \alpha}(e_{\gamma}-e_{\alpha})}, \qquad
|a_{\alpha}|^2=-\dfrac{\prod_{\gamma}(e_{\gamma}-E_{\alpha})}{\prod_{\gamma\neq \alpha}(E_{\gamma}-E_{\alpha})} .
\label{a_alpha}
\end{equation} 
With such values of $a_{\alpha}$ and $b_{\alpha}$ matrix  \eqref{tilde_a_b} is automatically unitary, $C\, C^{\dag}=1$.

Many different relations can be be derived for the above coefficients. In particular
\begin{equation}
\sum_{\beta}|b_{\beta}|^2=\sum_{\beta}|a_{\beta}|^2=\sum_{\beta}(E_{\beta}-e_{\beta}).
\end{equation}


\section{Probability distributions}\label{probability}

By construction  eigenvalues $e_{\alpha}$  and eigenfunctions $\Phi_1(\alpha)$ of matrix $G^{\beta}$  are distributed as in standard random matrix ensembles \cite{mehta}
\begin{equation}
P(\{ e_{\alpha}\} ,\, \{ r_{\alpha}\})\sim \prod_{\alpha<\gamma} |e_{\gamma}-e_{\alpha}|^{\beta} \,  \prod_{\alpha} r_{\alpha}^{\beta/2-1}\delta \left ( \sum_{\alpha} r_{\alpha}-1 \right )\,  \exp (-V(\{e_{\alpha}\} )).
\end{equation}
Here $r_{\alpha}$ is the modulus square of eigenfunction $\Phi_1(\alpha)$,  $r_{\alpha}=|\Phi_1(\alpha)|^2$ (of course, the same is valid for other components as well) 
and $V(\{e_{\alpha}\} )$ is a confinement term. For standard Gaussian ensembles \eqref{G_beta}
\begin{equation}
V(\{e_{\alpha}\} )= \frac{\beta}{4\sigma^2} \sum_{\alpha}  e_{\alpha}^2 .
\label{confinement}
\end{equation}
The mean density of matrix eigenvalues when $N\to\infty$ is given by the Wigner semicircle law (see e.g. \cite{mehta}) 
\begin{equation}
\rho_W(E)=\frac{1}{2\pi \sigma^2}\sqrt{4N\sigma^2-E^2}.
\label{wigner}
\end{equation}
For vector $v_j$ as in Eq.~\eqref{v}   coefficients  $b_{\alpha}\equiv \sum_j v_j \Phi_j(\alpha)$ are proportional to $\Psi_1(\alpha)$, $b_{\alpha}=\sqrt{Z} \Phi_1(\alpha)$.

According to Eq.~\eqref{quantisation_1} after the rank-one perturbation new eigenvalues are determined from the equation
\begin{equation}
Z \sum_{\beta}\frac{r_{\beta}}{E_{\alpha}-e_{\beta}}=1.
\end{equation}
Our purpose is to find the distribution of $\Psi_1(\alpha)$ elements. Using Eq.~\eqref{a_b} one gets
\begin{equation}
\Psi_1(\alpha)=\sum_{\beta} C_{\alpha \beta}\Phi_1(\beta)=\frac{1}{\sqrt{Z}}\sum_{\beta} C_{\alpha \beta} b_{\beta}=\frac{a_{\alpha}}{\sqrt{Z}}\sum_{\beta}\frac{|b_{\beta}|^2}{E_{\alpha}-e_{\beta}}=\frac{a_{\alpha}}{\sqrt{Z}}\ .
\end{equation}
Therefore the distribution of $\Psi_1(\alpha)$ coincides with distribution of $a_{\alpha}/\sqrt{Z}$. Denote 
\begin{equation}
z_{\alpha}=|\Psi_1(\alpha)|^2.
\end{equation}
From Eq.~\eqref{a_alpha}  it follows that 
\begin{equation}
z_{\alpha}=\dfrac{\prod_{\gamma}(E_{\alpha}-e_{\gamma})}{Z\prod_{\gamma\neq \alpha}(E_{\alpha}-E_{\gamma})}.
\end{equation}
The joint distribution of the old eigenvalues $e_{\alpha}$ and the new ones $E_{\alpha}$ without the confinement term \eqref{confinement}  (its absence is indicated below by a tilde) had been calculated in  Ref.~\cite{aleiner}
\begin{equation}
\tilde{P}(\{e_{\alpha}\},\{E_{\alpha}\})\sim \dfrac{\prod_{\gamma>\alpha} (e_{\gamma}-e_{\alpha})(E_{\gamma}-E_{\alpha})}{\prod_{\gamma,\alpha}|e_{\gamma}-E_{\alpha}|^{\beta/2-1}}\delta \left (\sum_{\alpha}(E_{\alpha}-e_{\alpha})-Z \right ).
\label{joint}
\end{equation} 
The joint distribution of new eigenvalues $E_{\alpha}$ and new eigenfunctions $z_{\alpha}\equiv |\Psi_1(\alpha)|^2$  is given by the expression
\begin{equation}
\tilde{P}(\{E_{\alpha}\}, \{z_{\alpha}\})=\prod_{\alpha}\delta\left (z_{\alpha}-\dfrac{\prod_{\gamma}(E_{\alpha}-e_{\gamma})}{Z \prod_{\gamma\neq \alpha}(E_{\alpha}-E_{\gamma})}\right )P(\{e_{\alpha}\},\{E_{\alpha}\}).
\end{equation}
One has $N$ variables $z_{\alpha}$ and $N$ variables $e_{\alpha}$. It is plain that 
\begin{equation}
\frac{\partial z_{\alpha}}{\partial e_{\beta}}=\frac{z_{\alpha}}{E_{\alpha}-e_{\beta}} \ .
\end{equation}
Using the Cauchy determinant one obtains
\begin{eqnarray}
\det \left ( \frac{\partial z_{\alpha}}{\partial e_{\beta}}\right )&=&\left (\prod_{\alpha} z_{\alpha} \right ) \det\left ( \frac{1}{E_{\alpha}-e_{\beta}}\right )=\left (\prod_{\alpha} z_{\alpha} \right ) 
\frac{\prod_{\alpha<\beta} (E_{\alpha}-E_{\beta})(e_{\beta}-e_{\alpha})}{\prod_{\alpha,\beta}(E_{\alpha}-e_{\beta})}\nonumber\\
&=&\frac{1}{Z^{N}}\prod_{\alpha<\beta} \frac{ (e_{\beta}-e_{\alpha})}{(E_{\alpha}-E_{\beta})}.
\end{eqnarray}
Finally 
\begin{equation}
\tilde{P}(\{E_{\alpha}\}, \{z_{\alpha}\})\sim \dfrac{\prod_{\gamma>\alpha}(E_{\gamma}-E_{\alpha})^2}{\prod_{\gamma,\alpha }|e_{\gamma}-E_{\alpha}|^{\beta/2-1}}\delta \left (\sum_{\alpha}(E_{\alpha}-e_{\alpha})-Z \right ).
\end{equation}
Because
\begin{equation}
\prod_{\alpha, \gamma}(E_{\alpha}-e_{\gamma})=\prod_{\alpha}z_{\alpha} (\prod_{\alpha<\gamma}(E_{\alpha}-E_{\gamma}))^2, \qquad 
\sum_{\alpha} z_{\alpha}=\frac{1}{Z}\sum_{\alpha}(E_{\alpha}-e_{\alpha})
\end{equation}
one comes to the conclusion that 
\begin{equation}
\tilde{P}(\{E_{\alpha}\}, \{z_{\alpha}\})\sim \prod_{\alpha<\gamma}|E_{\gamma}-E_{\alpha}|^{\beta}\prod_{\alpha}z_{\alpha}^{\beta/2-1}\delta \left (\sum_{\alpha}z_{\alpha}-1 \right ).
\end{equation}
It means that after a rank-one perturbation the joint distribution of new eigenvalues and eigenfunctions has the same form as the distribution  of initial quantities 
(cf. Eq.~\eqref{joint})
\begin{eqnarray}
&&\prod_{\alpha<\gamma} |e_{\gamma}-e_{\alpha}|^{\beta} \, \prod_{\alpha} r_{\alpha}^{\beta/2-1}\delta \left ( \sum_{\alpha} r_{\alpha}-1 \right )
\prod_{\alpha}\mathrm{d}e_{\alpha}\mathrm{d}r_{\alpha}=\nonumber\\
&=& \prod_{\alpha<\gamma}|E_{\gamma}-E_{\alpha}|^{\beta}\prod_{\alpha}z_{\alpha}^{\beta/2-1}\delta \left (\sum_{\alpha}z_{\alpha}-1 \right )
\prod_{\alpha}\mathrm{d}E_{\alpha}\mathrm{d}z_{\alpha}.
\end{eqnarray}
It seems that this key  identity has been overlooked in previous studies. This result could be anticipated without calculations when one notices that   Eq.~\eqref{def} can be written in the symmetric form $G_{ij}=M_{ij}-  v_i^{*}v_j$  
which interchanges $e_{\alpha}\leftrightarrow -E_{\alpha}$ but Eq.~\eqref{joint} is symmetric under this transformation.   

But this symmetry is valid only without the confinement term  \eqref{confinement}. Using the representation  $M_{ij}=\sum_{\alpha} E_{\alpha} \Psi_i(\alpha)\Psi_j^* (\alpha)$ and calculating $\mathrm{Tr}\, G^2=\mathrm{Tr}\, (M-Z\delta_{i1}\delta_{j1})^2 $ (or from the direct calculations as in Appendix of Ref.~\cite{integrable}) one concludes  that 
\begin{equation}
\sum_{\alpha}e_{\alpha}^2=\sum_{\alpha}E_{\alpha}^2-2Z\sum_{\alpha} E_{\alpha}z_{\alpha} + Z^2.
\end{equation}
Therefore the total  joint distribution of new eigenvalues $E_{\alpha}$ and new eigenvectors,  $z_{\alpha}\equiv |\Psi_1(\alpha)|^2$, is the following
\begin{eqnarray}
P(\{E_{\alpha}\}, \{z_{\alpha}\})&\sim& \prod_{\alpha<\beta}|E_{\beta}-E_{\alpha}|^{\beta}\prod_{\alpha}z_{\alpha}^{\beta/2-1} \delta \left (\sum_{\alpha}z_{\alpha}-1 \right )\nonumber\\
 &\times& \exp\left [ -\frac{\beta}{4\sigma^2}\left (\sum_{\alpha}E_{\alpha}^2-2Z \sum_{\alpha} E_{\alpha}z_{\alpha}\right )\right ]. 
\label{final}
\end{eqnarray}
 For the  initial distribution  eigenvalues and eigenvectors were independent but after a rank-one perturbation the distribution of eigenvectors depends on eigenvalues due to the term $\sum_{\alpha} E_{\alpha} z_{\alpha}$ in the exponent.


\section{Asymptotic calculations}\label{asymptotic}

Expression \eqref{final} is exact. In this Section only the most interesting case of large $N$ is considered though for $\beta=2$ certain analytical calculations 
are possible for finite $N$ \cite{brezin}. 


\subsection{Mean spectral  density} 

The simplest characteristic of the problem \eqref{M_G} is its mean spectral density. Though it is not necessary for further discussion,  its derivation is presented for completeness.    According to Ref.~\eqref{pastur} the averaged Green function for matrix $M$, $\bar{G}(E)=\frac{1}{N}\langle \mathrm{Tr}\, (E-M)^{-1}\rangle $  in the limit $N\to\infty$ obeys the equation
\begin{equation}
\bar{G}(E)=\frac{1}{N}\frac{1}{E-Z-\sigma^2 N \bar{G}(E)}+\frac{N-1}{N}\frac{1}{E-\sigma^2 N \bar{G}(E)}.
\label{pastur}
\end{equation}
Its solution in 2 lowest orders in $N^{-1}$ is
\begin{equation}
\bar{G}(E)=\bar{G}_0(E)\left (1+\frac{Z \bar{G}_0(E)} {N(1-Z\bar{G}_0(E))}\right )
\end{equation}
where $\bar{G}_0(E)$ is the mean Green function for standard RM ensembles which is determined from the equation
\begin{equation}
\bar{G}_0(E)=\frac{1}{E-\sigma^2 N \bar{G}_0(E)},\qquad \bar{G}_0(E)=\frac{E-\sqrt{E^2-4\sigma^2 N}}{2\sigma^2 N}\ .
\label{unperturbed_G}
\end{equation}
When $\kappa^2<1$ the mean level density is proportional to the imaginary part of $\bar{G}(E)$ which comes only from the square root in $\bar{G}_0(E)$ and exists when $-2\sigma \sqrt{N}<E<2\sigma \sqrt{N}$. The dominant term, of course,  is the Wigner semicircle law \eqref{wigner} (as it is evident from the intertwining of eigenvalues) but there is a correction term due to interaction. It is convenient to denote $E=2\sigma\sqrt{N}\cos \phi$ then the mean level density is
\begin{equation}
\rho(\phi)= \left (\frac{N}{2\pi}+\frac{2\kappa (2\cos \phi-\kappa)}{\pi(\kappa^2-2\kappa\cos \phi +1)}\right )\sin^2\phi .
\label{density_correction}
\end{equation}
For $\kappa^2<1$ the integral $\int_0^{\pi}\rho(\phi)\mathrm{d}\phi=N$ as it should be but for $\kappa^2>1$ this integral equals $N-1+1/\kappa^2$. The reason of it is well known and is related   with  the formation for $\kappa^2>1$ of one collective state well separated from the bulk with highest (when $Z>0$) or smallest (when $Z<0$) energy. This state manifests itself as an additional pole Eq.~\eqref{pastur} whose position, $E_c$  for large $N$ is determined by the zero of the denominator of the second term,  $1-Z \bar{G}_0(E_c)=0$ (which is simply the mean value of Eq.~\eqref{quantisation_1}).  When $\kappa^2>1$ the solution of this equation is 
\begin{equation}
E_c=\sigma \sqrt{N}\left ( \kappa+\frac{1}{\kappa} \right ).
\end{equation} 
In a vicinity of this energy $\bar{G}(E)\approx \bar{G}_0(E)+\delta G(E)$ where $\delta G(E)$ is determined from the first term of Eq.~\eqref{pastur}
\begin{equation}
\delta G(E)=\dfrac{1}{N\left (\frac{\kappa^2}{\kappa^2-1}(E-E_c)-\sigma^2 N\delta G(E)\right )}.
\end{equation} 
The imaginary part of this solution exists when $|E-E_c|< 2\sigma (1-\kappa^{-2})$ and it determines the additional level density
\begin{equation}
\delta \rho(E)= \frac{1}{\pi \sigma  a}\sqrt{ a^2-(E-E_c)^2}, \qquad a=2\sigma (1-\kappa^{-2}).
\end{equation}
When $\kappa^2>1$ the level density is non zero in two distinct regions. One corresponds to energy $|E|<2\sigma\sqrt{N}$ where the density is given by Eq.~\eqref{density_correction}. The integral of density over this region equals $N-1+k^{-2} $. The second part of the density is non-zero in the region $|E-E_c|<2\sigma (1-\kappa^{-2})$ and the integral over such interval is $1-k^{-2}$. The sum of the both regions equals $N$ as it should be.


\subsection{$\kappa^2<1$}

In the case $\kappa^2<1$  and  $N\gg1$ components $z_{\alpha}=|\Psi_1(\alpha)|^2 $ for all energies are of the order of $N^{-1}$ and the condition  $\sum_{\alpha}z_{\alpha}=1$  can be taken into account as usual by the introduction of Lagrange multiplier
\begin{equation}
\delta \left (\sum_{\alpha}z_{\alpha}-1 \right )\longrightarrow \exp \left (-\mu (\sum_{\alpha}z_{\alpha}-1)\right ).
\label{Lagrange}
\end{equation}
After this substitution  the probability distribution  \eqref{final} factores and different $z_{\alpha}$ become independent, each $z_{\alpha}\equiv z(E_{\alpha})$ being distributed as follows
\begin{equation}
P(z ,E)=\Big (\mu-\frac{Z}{2\sigma^2} E\Big)^{\beta/2}(\pi z)^{\beta/2-1} \exp\left ( -\Big (\mu-\frac{\beta Z}{2\sigma^2} E\Big)z\right ).
\label{modifiedPT}
\end{equation} 
The value of $\mu$ has to be calculated  from the requirement that  the mean value of $\sum_{\alpha}z_{\alpha}$ equals  $1$. It leads to 
\begin{equation}
\sum_{\alpha}  \int_0^{\infty}z\,P(z ,E_{\alpha} )\, \mathrm{d}z=1\quad \longrightarrow \quad \frac{\beta Z}{2}\sum_{\alpha}\frac{1}{\mu- \frac{\beta Z}{2\sigma^2}E_{\alpha}}=1.
\end{equation}
The sum in this equation can be expressed through  the mean unperturbed Green function, $\bar{G}_0(E)$, given by  Eq.~\eqref{unperturbed_G}. After simple algebra one finds that when $\kappa^2<1$
\begin{equation}
\mu=\beta N \frac{\kappa^2+1}{2}.
\label{mu_k}
\end{equation}
Consequently,  the eigenfunction distribution of perturbed problem has the same functional form as the PT distribution Eqs.~\eqref{pt_1}, \eqref{pt_2}
\begin{equation}
P_{\beta}(x)=\dfrac{1}{(2\pi x)^{1-\beta/2}(l(E))^{\beta/2}}\exp \left ( -\frac{\beta x}{2 l(E)}\right )
\label{PTmodified}
\end{equation}
but the mean value of $x=N|\Psi_1(E)|^2$  depends on the energy and the coupling constant
\begin{equation}
l(E)\equiv N \langle |\Psi_1(E)|^2\rangle = \left ( \kappa^2+1-\frac{\kappa }{\sigma\sqrt{N}} E \right )^{-1}.
\label{finalPT}
\end{equation} 


\subsection{  $\kappa^2>1$}

When $\kappa^2>1$ one collective state becomes separated from other levels. All states except the collective one  have  values of $z_{\alpha}$  of the order of $N^{-1}$ and their probability distributions are calculated as in the previous Section. The only difference is that their normalisation is 
\begin{equation}
\sum_{\alpha=1}^{N-1}z_{\alpha}=1-z_c
\end{equation}
where it is anticipated that $z_c$ for the collective state remains constant  when $N\to\infty$.

Performing the same calculations as above one  finds that the distribution of all  $z_{\alpha}$ (except the collective state) has the same form as  Eq.~\eqref{modifiedPT} but instead of \eqref{mu_k}  $\mu$ is given by the expression  (valid provided $|k|(1-z_c)<1$)
\begin{equation}
\mu(z_c)=\frac{\beta N}{2}\left ( \kappa^2(1-z_c)+\frac{1}{1-z_c}\right ) .
\label{mu_c}
\end{equation}
The knowledge of these distributions permits to find the distribution of the both $E\equiv E_c$ and $r\equiv z_c$ by integrating Eq.~\eqref{final} over $z_{\alpha}$
\begin{eqnarray}
& &P(E,r)\sim \prod_{\alpha}|E-E_{\alpha}|^{\beta} \left (\mu(r) -\frac{\beta \sqrt{N}\kappa}{2\sigma}E_{\alpha}\right )^{-\beta/2} \nonumber\\
&\times&\exp \left (-\frac{\beta}{4\sigma^2}E^2+\frac{\beta \sqrt{N}\kappa}{2\sigma} E r  -(1-r)\mu(r) \right )\sim \mathrm{e}^{-N \beta \,F(E ,r)}
\end{eqnarray} 
where the corresponding exponent is
\begin{eqnarray}
F(E,r)&=&-\frac{1}{N}\sum_{\alpha} \ln (E-E_{\alpha})+\frac{1}{4\sigma^2 N}E^2 +\frac{1}{2 N}\sum_{\alpha}  \ln \left (\nu(r)  -\frac{ \kappa }{2\sigma\sqrt{N}}E_{\alpha}\right )\nonumber\\
&-&\frac{\kappa }{2\sigma \sqrt{N}} E  r -(1-r) \nu(r),\qquad \nu(r)=\frac{1}{2}\left ( \kappa^2(1-r)+\frac{1}{1-r}\right ) .
\end{eqnarray}
In the limit $N\to\infty$ values of $E$ and $r$ are determined from the saddle point equations
\begin{eqnarray}
\frac{\partial F(E,r)}{\partial E} &= &-\frac{1}{N} \sum_{\alpha} \frac{1}{E-E_{\alpha}} +\frac{1}{2\sigma^2 N }E-  \frac{ \kappa}{2\sigma \sqrt{N}}  r=0, \\
\frac{\partial F(E,r)}{\partial r}   &=&\frac{1}{2 N}\sum_{\alpha} \frac{1}{\nu(r)-\frac{ \kappa}{2\sigma \sqrt{N} }E_{\alpha}} \frac{\partial \nu(r)}{\partial r}-\frac{ \kappa}{2\sigma \sqrt{N}} E-\frac{\partial [(1-r)\nu(r)]}{\partial r}=0.
\end{eqnarray}
As all arguments in sums over $E_{\alpha}$ are outside the spectrum, it is legitimate in the leading order to use instead of these sums their mean value \eqref{unperturbed_G}. It leads to  the system of equations
\begin{equation}
 \sqrt{E^2-4\sigma^2 N}-\sigma \sqrt{N}\kappa r=0, \qquad 
(1-r)\Big (-\kappa^2+\frac{1}{(1-r)^2}\Big)-\frac{\kappa}{\sigma\sqrt{N}} E +2\kappa^2 (1-r)=0
\end{equation}
whose solution gives the following values for saddles  
\begin{equation}
E_c=\sigma\sqrt{N}\Big ( \kappa+\frac{1}{\kappa} \Big),\qquad r_c=1-\frac{1}{\kappa^2}.
\end{equation} 
Substituting $r_c=1-\kappa^{-2}$ to Eq.~\eqref{mu_c} one concludes that for $\kappa^2>1$ the modified PT distribution has exactly the same form as for $\kappa^2<1$ and is given by Eq.~\eqref{PTmodified}. 

\section{Large window distribution}\label{large}

The obtained simple Gaussian-like formulae  correspond  to distribution of eigenfunctions with fixed energy or, more precisely, in  small energy intervals $|\delta E| \ll \sigma \sqrt{N}$. For practical reasons it is important to know the distribution of eigenfunctions $z_{\alpha}=N |\Psi_1(\alpha)|^2$ whose energies $E_{\alpha}$ are in a finite interval $E_1<E_{\alpha}< E_2$. The above results stipule that the moments of the resulting distribution have to be calculated from the expression
\begin{equation}
\langle z_{\alpha}^q\rangle_{[E_2,E_1]} =\frac{c_{\beta}(q)}{\delta N}\int_{E_1}^{E_2}\rho_W(E)\left ( \kappa^2+1-\frac{\kappa }{\sigma\sqrt{N}} E \right )^{-q}\mathrm{d}E,\qquad \delta N=\int_{E_1}^{E_2}\rho_W(E)\mathrm{d}E
\label{moments_large_intervals}
\end{equation}
where $\rho_W(E)$ is the Wigner spectral density  \eqref{wigner} and $c_{\beta}(q)$ are the Gaussian moments
\begin{equation}
c_1(q)=\frac{2^q \Gamma(q+1/2)}{\sqrt{\pi}},\qquad c_2(q)=\Gamma(q+1).
\end{equation}

Similarly, the full distribution in a finite interval is the weighted integral of Eq.~\eqref{PTmodified} 
\begin{equation}
\mathcal{P}_{\beta}(x)=\frac{1}{\delta N}\int_{E_1}^{E_2} \dfrac{\rho_W(E) }{(2\pi x)^{1-\beta/2}(l(E))^{\beta/2}}\exp \left ( -\frac{\beta x}{2l(E)}\right ) \mathrm{d}E, \quad l(E)=\frac{1}{\kappa^2+1-\frac{\kappa }{\sigma\sqrt{N}} E }.
\label{distribution_large}
\end{equation}
In particular, when all states (except the collective one, if any) are taken into account $E_1=-2\sigma \sqrt{N}$ and $E_2=2\sigma \sqrt{N}$. Straightforward calculations show that for $\beta=1$ 
\begin{equation}
\mathcal{P}_{1}(x)=P_1(x)\, \mathcal{F}_1(x), \quad  \mathcal{F}_1(x)=\frac{2}{\pi}\int_0^{\pi}\mathrm{d}\phi \sin^2\phi\sqrt{ \kappa^2+1-2\kappa \cos \phi }\, \mathrm{e}^{ -\tfrac{1}{2} (\kappa^2-2\kappa \cos \phi )x },
\end{equation}
and   for $\beta=2$
\begin{equation}
\mathcal{P}_{2}(x)=P_2(x)\, \mathcal{F}_2(x), \quad  \mathcal{F}_2(x)=\frac{I_1(2\kappa x)}{\kappa x}\,\mathrm{e}^{-\kappa^2 x}.
\end{equation}
Here $P_{1,2}(x)$ are the normalised  PT distributions \eqref{pt_1}, \eqref{pt_2} with $l=1$ and $I_1(x)$ is the modified Bessel function. 

For small values of $\kappa$ one can use the series expansion in power of $\kappa$
\begin{eqnarray}
\mathcal{F}_1(x)&= & 1+\frac{\kappa^2}{8}(x^2-6x+3)+\frac{\kappa^4}{192}(x^4-16x^3+54x^2-24 x-3)+\mathcal{O}(\kappa^6),\\
\mathcal{F}_2(x)&= & 1+\frac{\kappa^2}{2}(x^2-2 x)+\frac{\kappa^4}{12} (x^4-6x^3+6x^2)+\mathcal{O}(\kappa^6).
\end{eqnarray}
These series give good results for $\kappa<.5$ and $x<10$. In principle  one can also approximate the above expression by series in the Hermite polynomials as for nearly Gaussian distributions but simplicity of the results makes approximate formulae unnecessary. 

\begin{figure}
\begin{center}
\includegraphics[width=.7\linewidth]{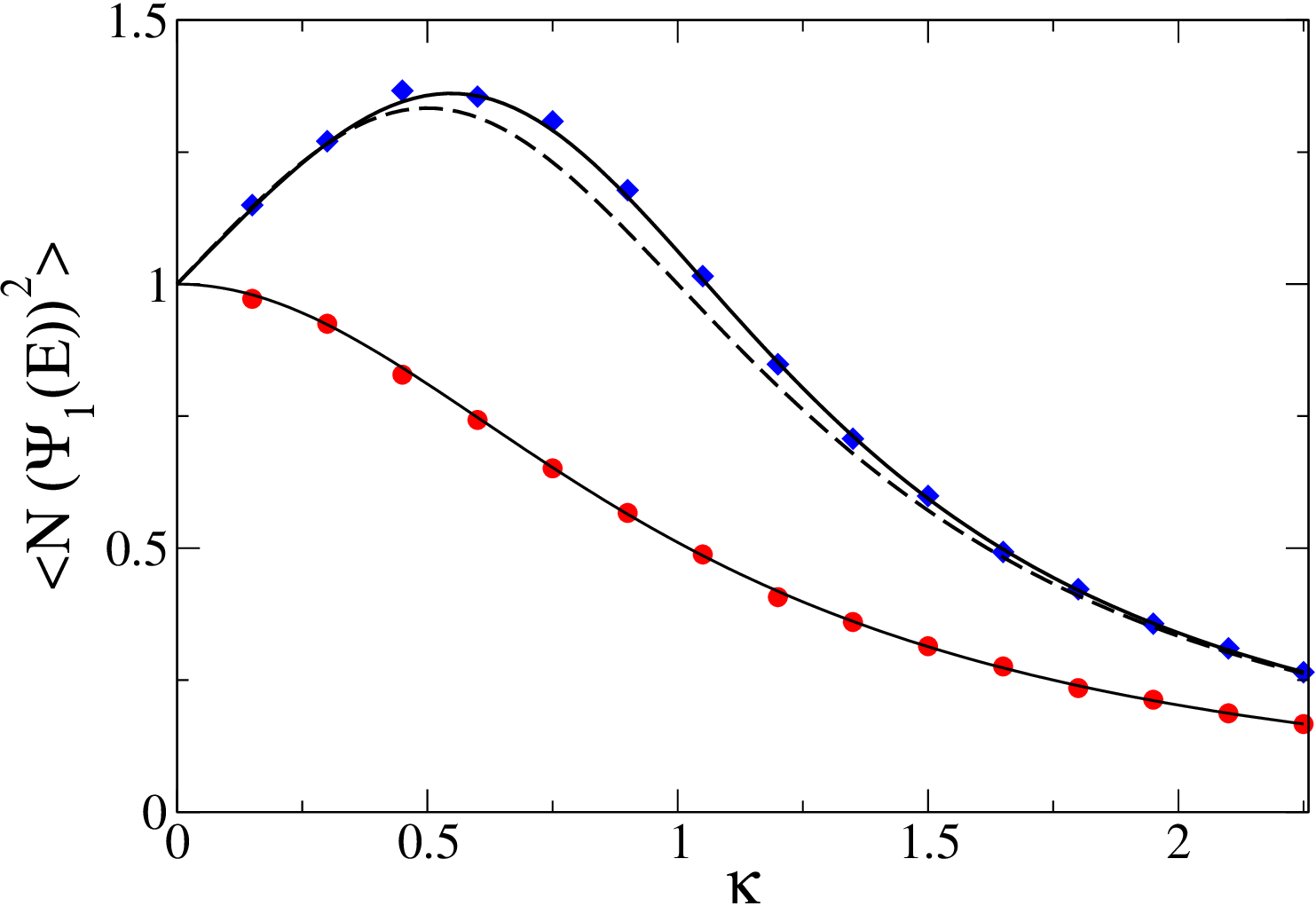}
\end{center}
\caption{Mean values of $N\langle (\Psi_1(E))^2\rangle$ for different $\kappa$. Lower (red) circles are mean values for states with energies in the interval $[-\sqrt{N}/2,\sqrt{N}/2]$. Upper (blue) diamonds are the same but for energies in the interval 
$[\sqrt{N}/2,3\sqrt{N}/2]$. Solid black lines represent theoretical predictions for these quantities given by Eq.~\eqref{moments_large_intervals}. Dashed black line is the prediction Eq.~\eqref{finalPT} for mean value in a small interval. In calculations $N=1000$ and each point is averaged over $50$ random realisations.}
\label{fig_1}
\end{figure}

To illustrate the obtained results, numerical diagonalization of matrices  \eqref{M_G} with dimension $N=1000$ and $\sigma=1$ has been performed. In Fig.~\ref{fig_1} average values of variable $x=\sqrt{N}\Psi_1(E)$  for real symmetric matrices ($\beta=1$)  calculated numerically are presented for different coupling constants $\kappa$. States with energies in two large intervals $I_1=[-\sqrt{N}/2,\sqrt{N}/2]$ and $I_2=[\sqrt{N}/2,3\sqrt{N}/2]$ were chosen. For each value of $\kappa$ one calculates  the average over all states in each interval and for 50 realisations of random matrices.  The numerical results are compared with two predictions  for this quantity, Eq.~\eqref{finalPT} valid for small intervals and more accurate 
Eq.~\eqref{moments_large_intervals} for large intervals. In   Eq.~\eqref{finalPT}  the interval centre is used as $E$. For interval $I_1$ these two predictions are practically indistinguishable but, as expected, for interval $I_2$ the second formula works better. It is clearly seen that  agreement between analytical formulae and  numerics is excellent.  

\begin{figure}
\begin{center}
\includegraphics[width=.49\linewidth]{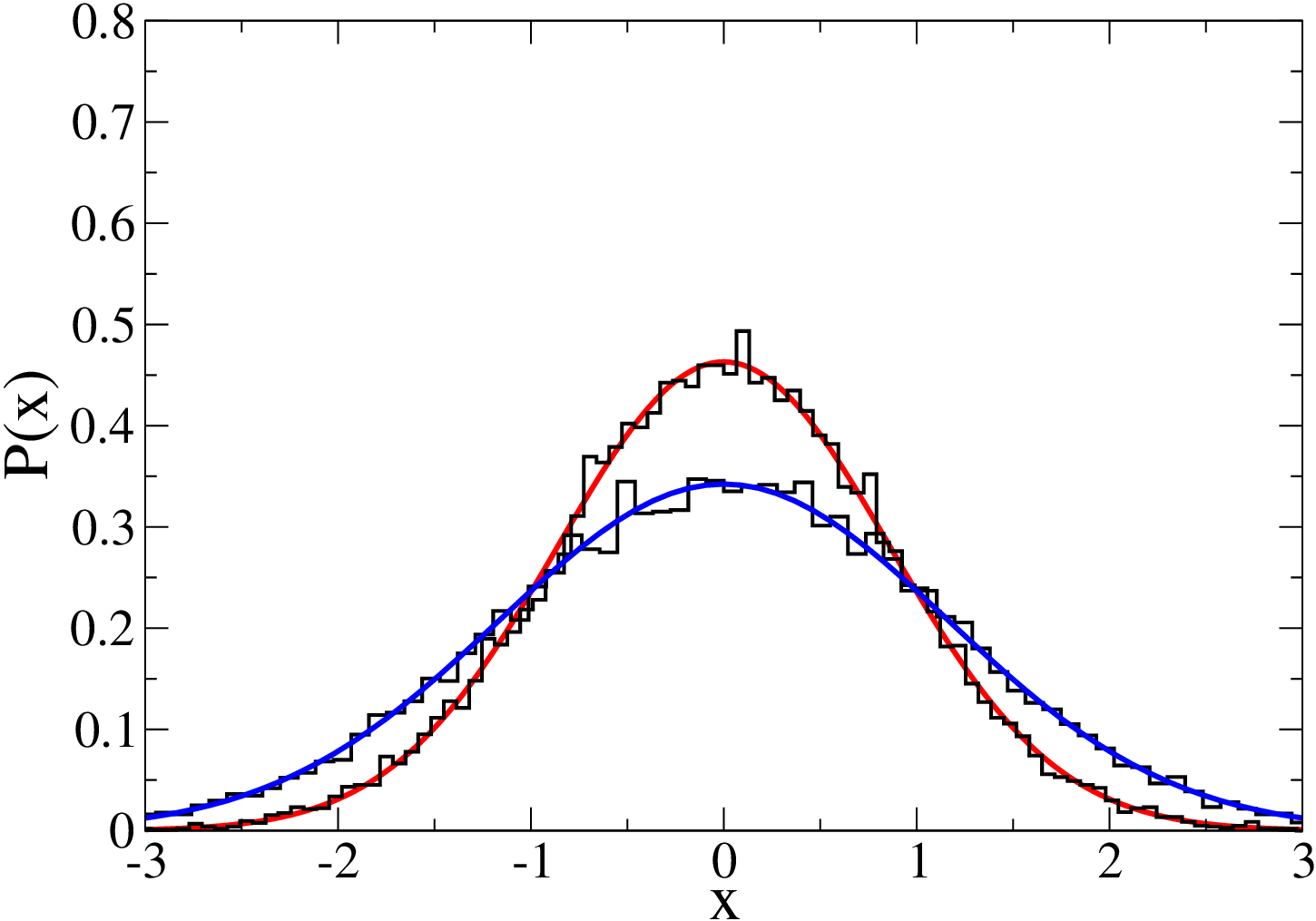}\hfill
\includegraphics[width=.49\linewidth]{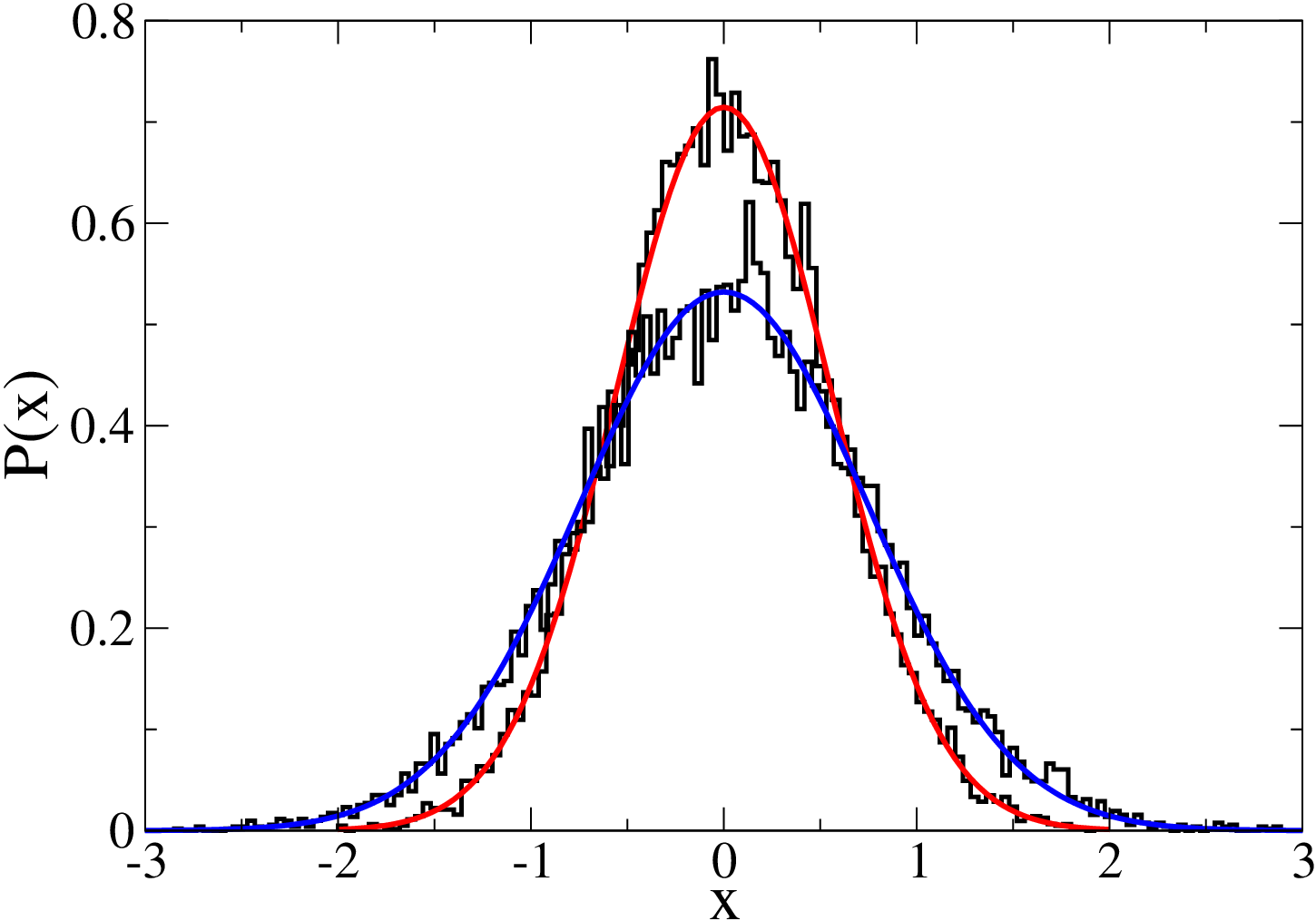}
\end{center}
\caption{Left: Distribution of $x=\sqrt{N} \Psi_1(E)$ for $\beta=1$ and $\kappa=0.6$ for states with energies in  $[-\sqrt{N}/2,\sqrt{N}/2]$ (lower curve) and $[\sqrt{N}/2,3\sqrt{N}/2]$ (upper curve).  Right: the same but  for $\kappa=1.5$. Solid lines are the Gaussian fits whose parameter agree well with theoretical predictions.}
\label{fig_2}
\end{figure}

\begin{figure}
\begin{center}
\includegraphics[width=.49\linewidth]{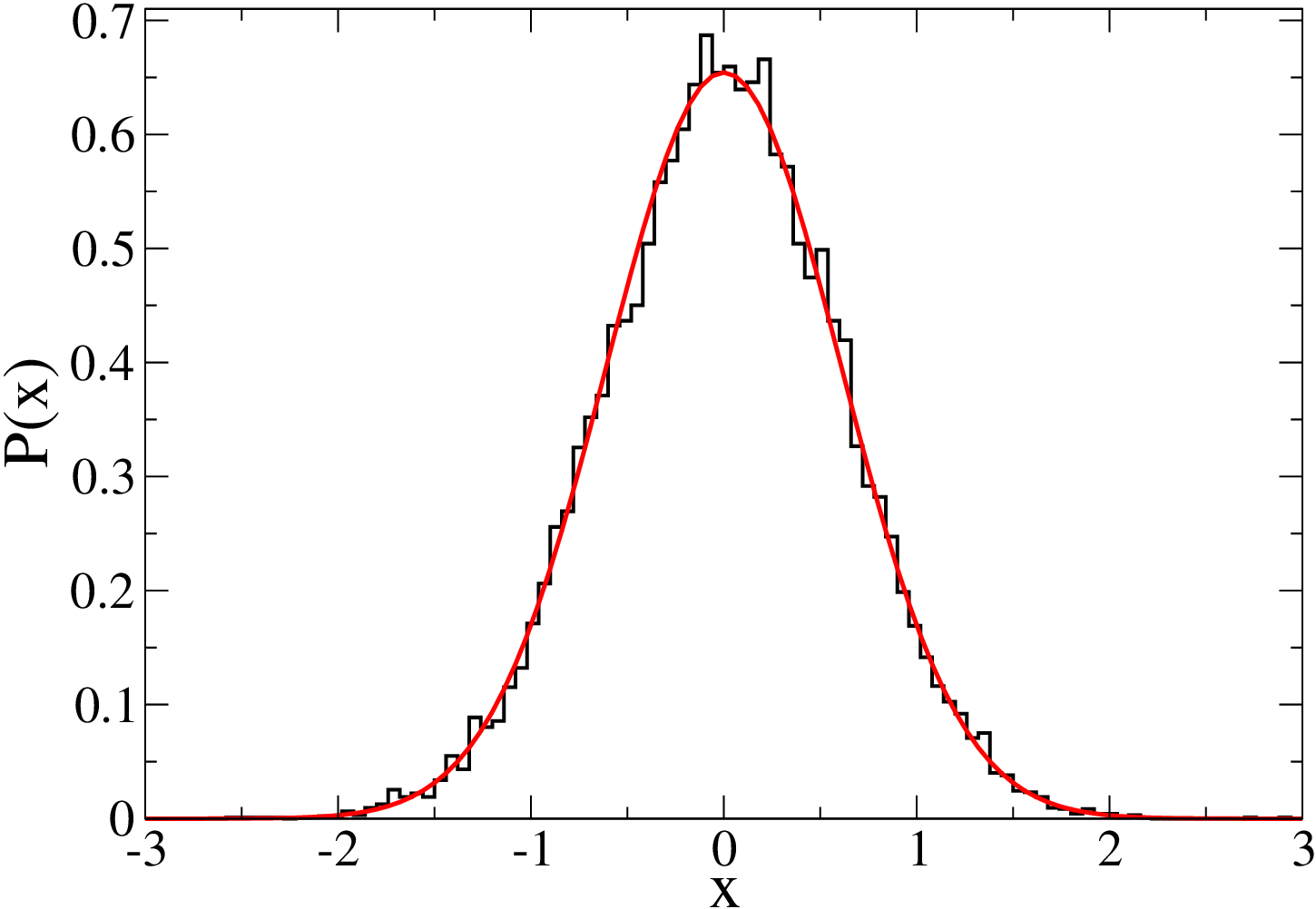}\hfill
\includegraphics[width=.49\linewidth]{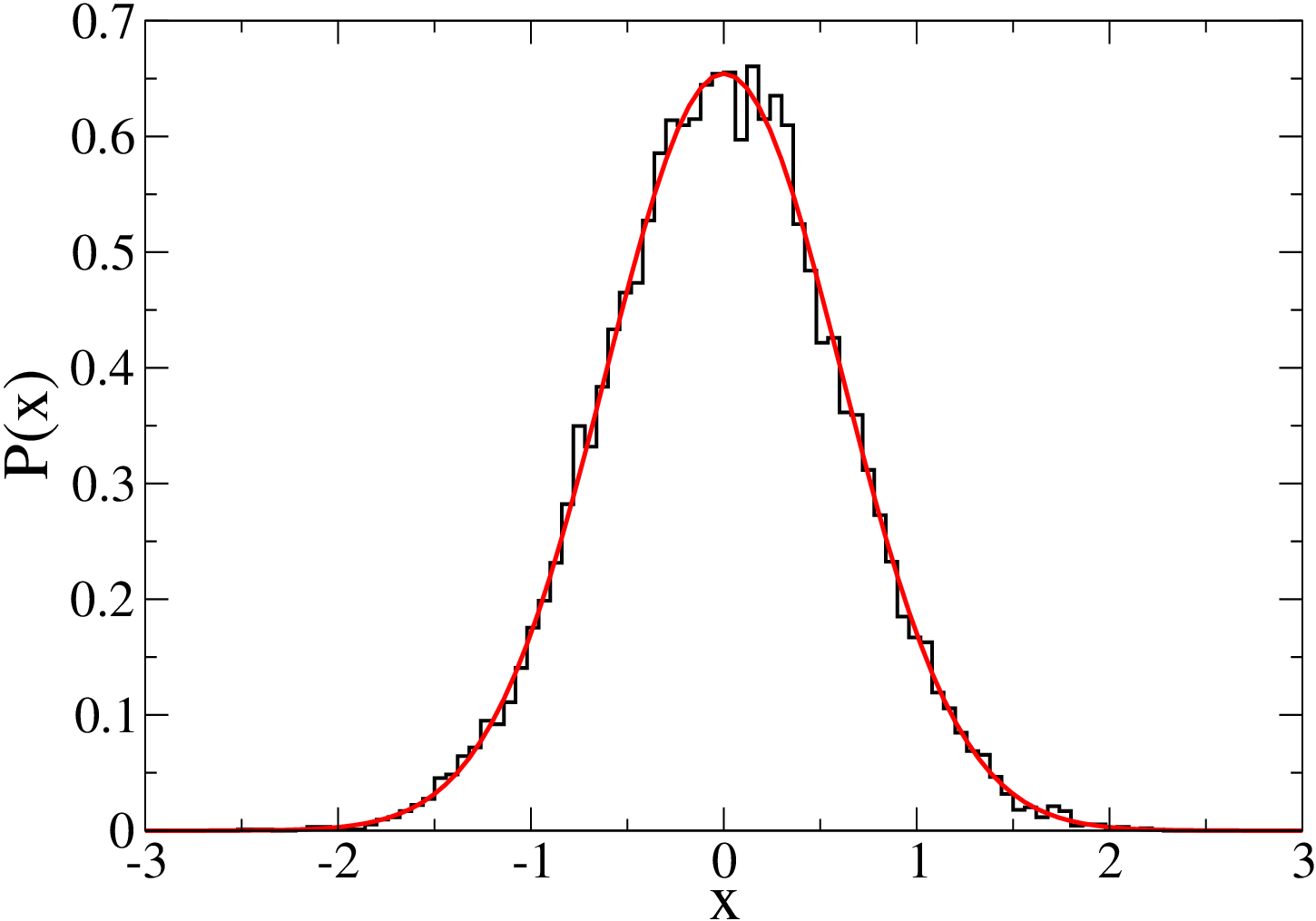}
\end{center}
\caption{Left: Distribution of  real parts of eigenfunctions,  $x=\sqrt{N}\, \mathrm{Re}\,  \Psi_1(E)$  for complex Hermitian random matrices with  $\kappa=0.6$  for  energies in  interval $[-\sqrt{N}/2,\sqrt{N}/2]$. Right: The same but for imaginary parts,  $x=\sqrt{N}\, \mathrm{Im}\,  \Psi_1(E)$. Solid red lines are Gaussian fits. }
\label{fig_3}
\end{figure}

\begin{figure}
\begin{center}
\includegraphics[width=.49\linewidth]{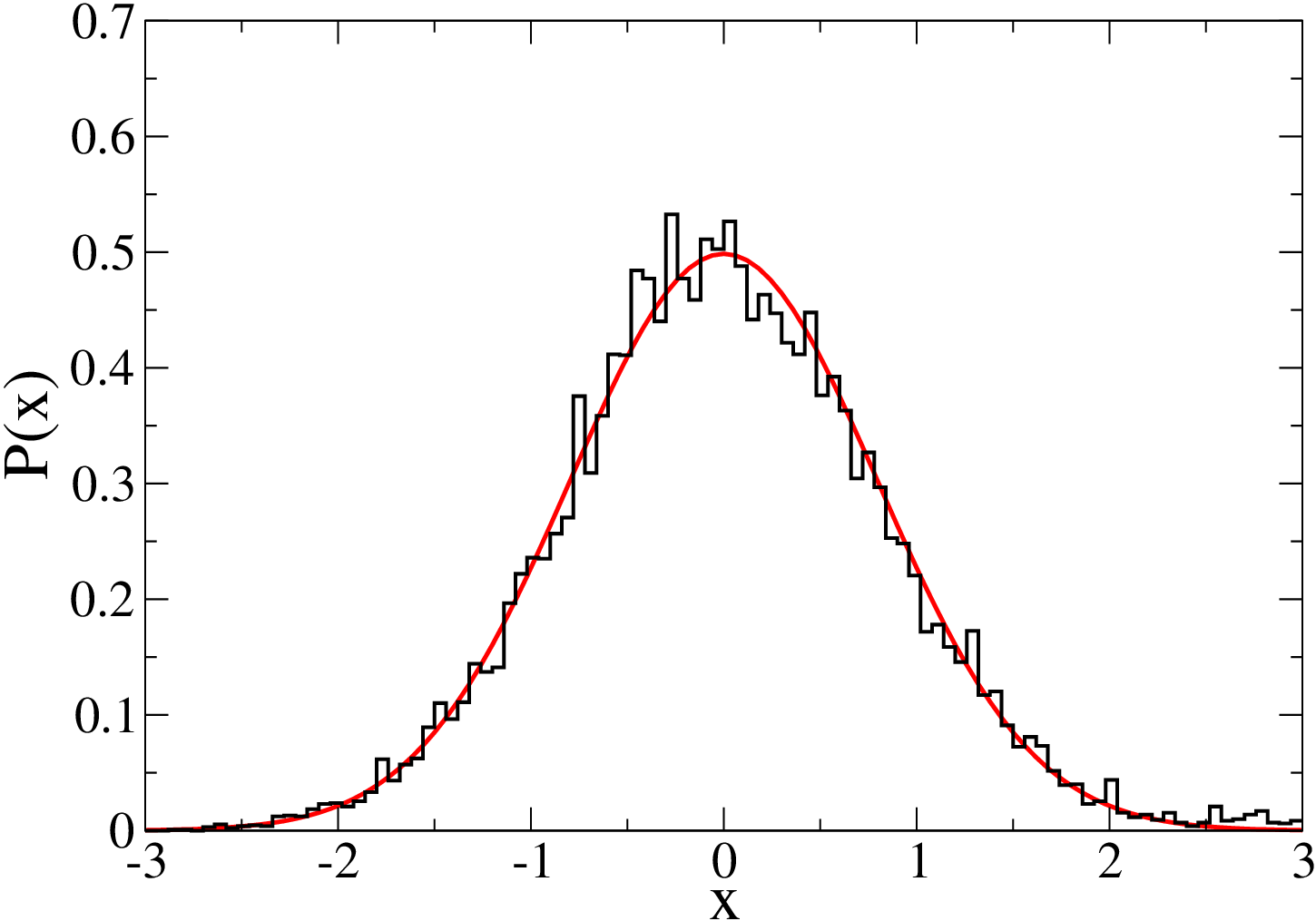}\hfill
\includegraphics[width=.49\linewidth]{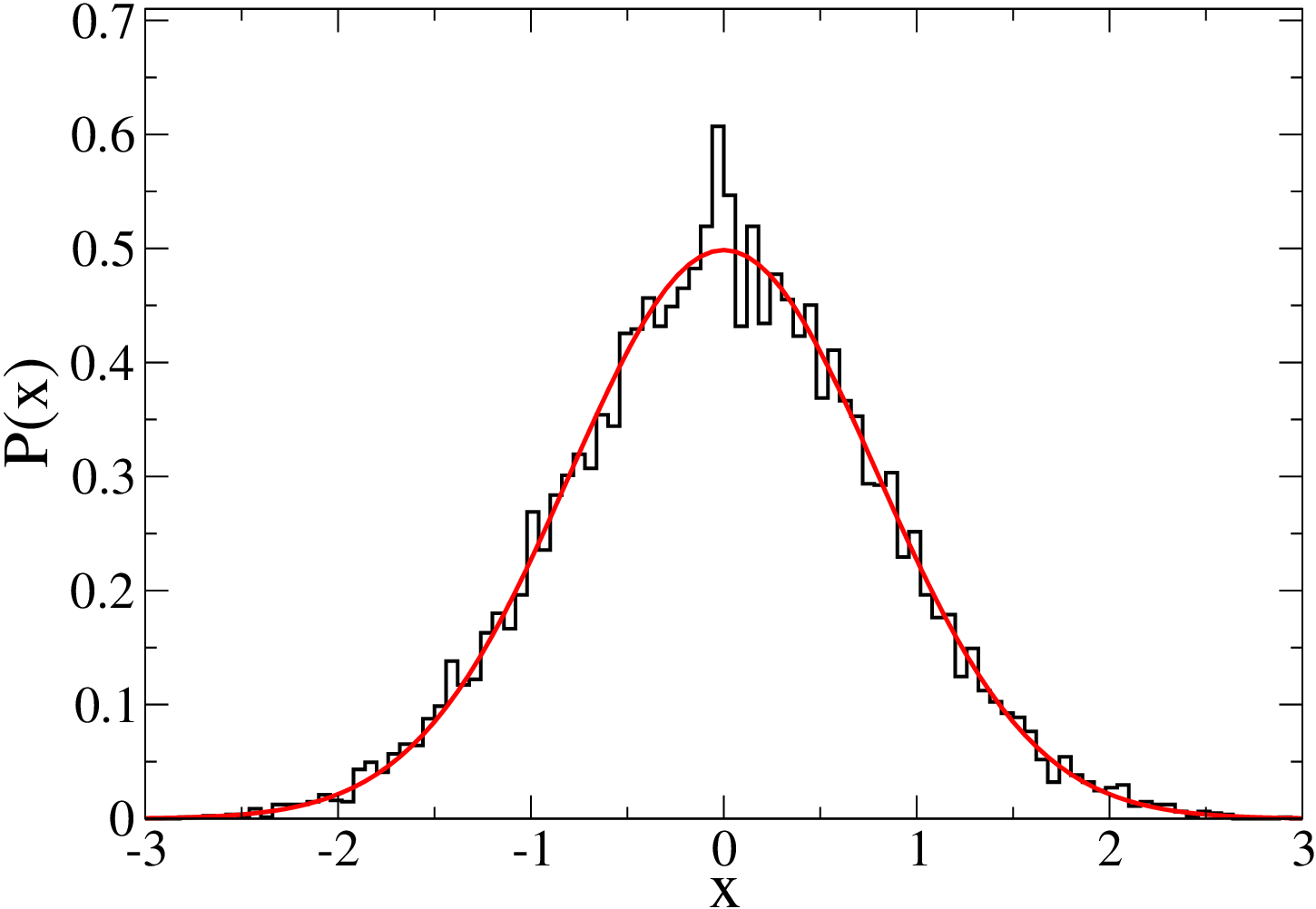}
\end{center}
\caption{The same as in Fig.~\ref{fig_3} but  for states with energies in  $[\sqrt{N}/2,3\sqrt{N}/2]$.}
\label{fig_4}
\end{figure}

To check the distribution shape the histograms of $x=\sqrt{N} \Psi_1(E)$ for real symmetric matrices ($\beta=1$ case)  are presented in Fig.~\ref{fig_2}. The PT representation \eqref{pt_1}  due to singularity is less sensitive to small deviations than the distribution of $\sqrt{N} \Psi_1(E_{\alpha})$.
In Fig.~\ref{fig_3} the distributions of the real and imaginary parts of $\sqrt{N} \Psi_1(E)$ are presented for complex Hermitian matrices (with $\beta=2$) with $\kappa=0.6$ when the energy are selected in the interval $I_1=[-\sqrt{N}/2,\sqrt{N}/2]$ and  in Fig.~\ref{fig_4} the same quantities but for the interval $I_2=[\sqrt{N}/2,3\sqrt{N}/2]$ are plotted. In all cases Gaussian fit perfectly fit the data and its width agrees well with predictions.

\section{Conclusion}\label{conclusion}

The standard PT distribution stipules that eigenvectors of large random matrix are independent identically distributed random variables whose probability density  for 
$x=N|\Psi|^2$ are given by Eqs.~\eqref{pt_1} and \eqref{pt_2} with $l=1$. This distribution is universal and has no adjustable parameters.

The main conclusion of the  paper is  that when ensemble of standard random matrices with Gaussian distribution \eqref{G_beta} is perturbed by a rank-one perturbation $Z\delta_{1\, i}\delta_{1\,j}$,  the distribution of $x=N|\Psi_1(E)|^2$ has the same functional form as the PT distribution but parameter $l$ entered  Eqs.~\eqref{pt_1} and \eqref{pt_2}  is not an universal constant but depends on energy $E$ and the coupling constant $\kappa=Z/(\sigma \sqrt{N})$
\begin{equation}
l(E) \equiv \langle  N\, |\Psi_1(E)|^2 \rangle= \left ( \kappa^2+1-\frac{\kappa }{\sigma \sqrt{N}} E \right )^{-1}.
\end{equation}
This expression is valid for  $\kappa^2<1$ as well as for $\kappa^2>1$. In the latter case there exists one collective state whose mean energy is $E_c=\sigma \sqrt{N}(\kappa+\kappa^{-1})$. The modulus square of the corresponding eigenvector is not of the order of $N^{-1}$ as all others components but is independent on $N$. More precisely, $\langle |\Psi_c|^2\rangle =1-\kappa^{-2}$. The Gaussian character of the obtained distributions is valid only for eigenfunctions in small energy intervals. When all eigenfunctions from a large energy interval are considered their distribution is not Gaussian but is given by an integral over Gaussian functions \eqref{distribution_large}.  In the limit $N\to\infty$ all other components of eigenfunctions (except $\Psi_1(E)$)  remain distributed according to the usual PT distribution.  The considered model is a rare example of non-invariant matrix models whose eigenfunction distribution  is derived analytically. 

The important  difference between the calculated distribution  and the PT one is that the latter is universal but the former is not. The interaction couples eigenfunctions with eigenenergies and forces the distribution to depend on coupling constant, state energy, and the form of confinement potential. For different resonances (e.g. for different nuclei) these quantities may and will be different. The simplest way to check these idea experimentally is to fit a width distribution for a particular resonance in a small energy window to the PT formula~\eqref{pt_1} and find the corresponding $l(E)$ from the fit. The absence of a priori restrictions on the dependence $l(E)$ on energy (it may have e.g. a power dependence or singularities) makes this approach quite flexible to describe various experimental data. Taking into account  together different resonances with different energies as it is often done to increase statistics is not a sensitive way to investigate this phenomenon.  

An interesting feature of the considered model~\eqref{M_G}  is that the introduction of rank-one interaction does not change local spectral statistics \cite{brezin}, \cite{bogomolny}.  Irrespective of the interaction strength statistical properties of eigenvalues at the scale of mean level density remain the same as for non-perturbed matrix (i.e. GOE or GUE) as it can be seen from the joint eigenvalue distribution \eqref{final}. Experimentally it was observed that the $\Delta_3$ statistics of nuclear resonances at small distances does agree well with RM prediction but  becomes to deviates from it  at distances of the order of 40-70 mean level spacings \cite{koehler_2}.  Large distance deviations from RM formulae are typical  for dynamical systems \cite{berry} and, in general, is not an argument against applicability  of  RM theory. 

\begin{acknowledgments}
The author is grateful to  D. Savin for pointing out Ref.~\cite{volya} and to ICTP, Trieste, for hospitality during the visit where the paper has been written.   
\end{acknowledgments}

\end{document}